\documentclass[final,nobibnotes,floatfix,
amsmath,amssymb,
aps,prb,showpacs,
twocolumn,
letterpaper,
longbibliography,
lengthcheck
]{elsarticle}

\usepackage{graphicx}
\usepackage{rotating}
\usepackage{amsmath}
\usepackage{bbm}
\usepackage{subfigure}
\usepackage[usenames,dvipsnames]{color}
\usepackage{natbib}

\renewcommand{\v}[1]{{\bf #1}}

\newcommand{\be}{\begin{equation}}
\newcommand{\ee}{\end{equation}}
\newcommand{\bea}{\begin{eqnarray}}
\newcommand{\eea}{\end{eqnarray}}

\graphicspath{
  {figures/pdf/}
  {figures/eps/}
}

\begin{document}

\title{Time-dependent density-functional and reduced density-matrix
       methods for few electrons: Exact versus adiabatic
       approximations}

\author[nanobio,etsf]{N.~Helbig\corref{cor1}}
\ead{nehelbig@gmail.com}
\author[nanobio,etsf]{J.I.~Fuks}
\author[nanobio,iker,etsf]{I.V.~Tokatly}
\author[fhi,etsf]{H.~Appel}
\author[halle,etsf]{E.K.U.~Gross}
\author[nanobio,fhi,etsf]{A.~Rubio}

\address[nanobio]{Nano-Bio Spectroscopy group, Dpto.~F\'isica de
  Materiales, Universidad del Pa\'is Vasco, Centro de F\'isica de
  Materiales CSIC-UPV/EHU-MPC and DIPC, Av.~Tolosa 72, E-20018 San
  Sebasti\'an, Spain}
\address[iker]{IKERBASQUE, Basque Foundation for Science, E-48011 Bilbao, Spain}
\address[fhi]{Fritz-Haber-Institut der Max-Planck-Gesellschaft,
Faradayweg 4-6, D-14195 Berlin, Germany}
\address[halle]{Max-Planck-Institut f\"{u}r Mikrostrukturphysik,
Weinberg 2, D-06120 Halle, Germany}
\address[etsf]{European Theoretical Spectroscopy Facility}

\cortext[cor1]{Corresponding author:}

\begin{abstract}
To address the impact of electron correlations in the linear and
non-linear response regimes of interacting many-electron systems
exposed to time-dependent external fields, we study
one-dimensional (1D) systems where the interacting problem is
solved exactly by exploiting the mapping of the 1D $N$-electron
problem onto an $N$-dimensional single electron problem. We
analyze the performance of the recently derived 1D local density
approximation as well as the exact-exchange orbital
functional for those systems. We show that the interaction
with an external resonant laser field shows Rabi oscillations
which are detuned due to the lack of memory in adiabatic approximations.
To investigate situations where static
correlations play a role, we consider the time-evolution of the
natural occupation numbers associated to the reduced one-body
density matrix. Those studies shed light on the non-locality
and time-dependence of the  exchange and correlation functionals
in time-dependent density and density-matrix functional theories.
\end{abstract}

\date{\today}

\maketitle

\section{Introduction}

Since its invention in 1984 time-dependent density-functional
theory (TDDFT) has become one of the major tools for describing
time-dependent phenomena of electronic systems
\cite{RG1984,TDDFT2006}. Despite its success, several important
questions remain open. A prominent example are double excitations \cite{EGCM2011},
which cannot be described with adiabatic approximations to the
exchange-correlation (xc) kernel \cite{MZCB2004}. Other examples
include the description of memory \cite{MBW2002}, charge-transfer
excitations \cite{DH2004}, Rabi oscillations \cite{BauerRugg}, and
population control \cite{RCWRG2008,BWG2005}. Also, the
construction of functionals for certain observables can be
problematic, like e.g.~double-ionization in strong laser fields
where one strategy rests on expressing the pair-correlation
function as a functional of the time-dependent density
\cite{PG1999}.

In many cases, there is little knowledge about how the dynamics of
the many-body system interacting with an arbitrary external
time-dependent field is mapped onto the non-interacting
(time-dependent) Kohn-Sham system. Here, one-dimensional systems
can provide insight since these systems can be exactly
diagonalized and subsequently propagated in time for a small
number of electrons. We provide insight into the limitations of
adiabatic functionals, especially for describing non-linear
electron dynamics exemplified by the case of Rabi oscillations.

This article is organized as follows, we first highlight the exact
mapping of a many-electron system onto an $N$-dimensional one-electron
problem and the selection of proper fermionic solutions. Then, we
discuss the recently developed one-dimensional local density
approximation (LDA) and its performance for calculating linear and
non-linear response. We use the LDA as well as exact exchange (EXX) to
investigate the description of double excitations and Rabi
oscillations with adiabatic approximations. We then change from TDDFT
to reduced density-matrix functional theory, where we discuss under
which conditions adiabatic approximations provide a valid description.
We conclude the paper with a short summary and perspectives.

\begin{figure*}
\begin{center}
\includegraphics{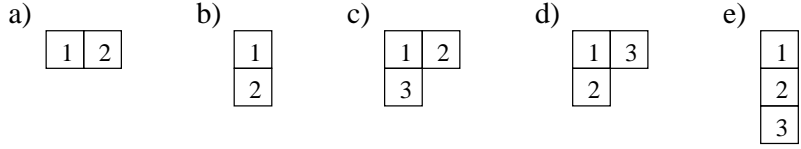}
\end{center}
\caption{\label{fig:young}
Possible standard Young diagrams for the spatial part of the wave
function for two [figures a) and b)] and three [figures c) to e)]
electrons. There are maximally two columns in each diagram, one for
each spin direction. Figures a) and b) correspond to the two electron
singlet and triplet, respectively. For diagram c) the wave function is
symmetrized for particles 1 and 2 and antisymmetrized for 1 and 3,
while for diagram d) the symmetrization is with respect to particles 1
and 3 and the antisymmetrization with respect to 1 and 2. For diagram
e) the wave function is antisymmetrized with respect to the
interchange of any two particles.}
\end{figure*}

\section{One-dimensional model systems}\label{sec:1D}

The Hamiltonian for $N$ electrons moving in a general, possibly
time dependent, external potential $v_{\rm ext}$ in one spatial dimension reads
\be\label{eq:1dham}
H=\sum_{j=1}^N \left[ -\frac{d^2}{2 dx_j^2}+v_{\rm ext}(x_j,t)\right]+\frac{1}{2}
\sum_{\stackrel{\scriptstyle j,k=1}{j\neq k}}^N v_{\rm int}(x_j, x_k),
\ee
where $v_{\rm int}$ describes the electron-electron interaction
(atomic units $e=m=\hbar=1$ are used throughout this paper). In one
spatial dimension the singularity of the ordinary Coulomb interaction
prevents electrons from passing the position of the singularity, both
in the attractive and repulsive case. In order to avoid this
unphysical behavior of the full Coulomb interaction we employ the so
called soft-Coulomb interaction
\be\label{eq:softC}
v_{\rm soft-C}(x_1, x_2) = \frac{q_1q_2}{\sqrt{a^2+(x_1-x_2)^2}}
\ee
instead \cite{SE1991}. Here, $q_1$ and $q_2$ describe the charges of
the particles while $a$ is the usual softening parameter. We use $a=1$
for all our calculations. Mathematically, it is straightforward to
show that the Hamiltonian (\ref{eq:1dham}) is equivalent to a
Hamiltonian for a single particle in $N$ dimensions moving in an
external potential
\be
v_{\rm Ndim}(x_1...x_N)=\sum_{j=1}^N v_{\rm ext}(x_j)+\frac{1}{2}
\sum_{\stackrel{\scriptstyle j,k=1}{j\neq k}}^N v_{\rm int}(x_j, x_k)
\label{eq:vdimN}
\ee
consisting of all the contributions from $v_{\rm ext}$ and $v_{\rm
  int}$. The corresponding Schr\"o\-dinger equation can, hence, be
solved by any code which is able to treat non-interacting particles in
the correct number of dimensions in an arbitrary external potential.

Due to the Hamiltonian being symmetric under particle interchange,
$x_j\leftrightarrow x_k$, the solutions of the Schr\"odinger equation
can be classified according to irreducible representations of the
permutation group.  For the simplest case of two interacting electrons
both the symmetric and antisymmetric solutions are valid corresponding
to the singlet and triplet spin configurations, respectively. For more
than two electrons one needs to separately ensure that the spatial
wave function is a solution to the $N$-electron problem. For example,
a totally symmetric spatial wave function is a correct solution for a
single particle in $N$ dimensions, however, for $N>2$ there is no
corresponding spin function such that the total wave function has the
required antisymmetry to be a solution of the $N$ particle problem in
1D. We solve this problem by symmetrizing the solutions according to
all possible fermionic Young diagrams for the given particle number
$N$ \cite{LL1977}.  Fig.~\ref{fig:young} shows all possible standard
Young diagrams for the spatial part of the wave function for two and
three electrons. As the spin of the electron is $1/2$, the Young
diagrams for the spin part can maximally have two rows, one for each
spin direction. The Young diagrams for the spatial part of the wave
function are then given as the transpose of the respective spin
diagram and, hence, have at most two columns. For two electrons there
exist two diagrams corresponding to the singlet (Fig.~\ref{fig:young}
a) and triplet (Fig.~\ref{fig:young} b) configurations. For three
electrons, there exist two diagrams with two electrons in one spin
channel and the remaining electron in the other channel
(Fig.~\ref{fig:young} c and d) and one diagram with all electrons
having the same spin (Fig.~\ref{fig:young} e).

In practice, we solve the Schr\"odinger equation in $N$ dimensions and
then symmetrize each solution according to the Young diagrams for the
given particle number. If none of the Young diagrams yields a
non-vanishing solution after symmetrization the state does not
describe a solution for spin-$1/2$ particles and is discarded. If a
state yields a non-vanishing contribution for a given diagram the
appropriately symmetrized state is normalized and used in further
calculations.

The solution of higher dimensional problems within these symmetry
restrictions has been implemented in the {\tt octopus} computer
program \cite{MCBR2003,CAORALMGR2006}. The lowest energy solution is
found to be purely symmetric and is, therefore, for $N>2$,
discarded. With increasing number of electrons we also observe an
increasing number of states which do not satisfy the fermionic
symmetry requirements.

\section{Local density approximation}\label{sec:lda}

The local density approximation for electrons interacting in one
spatial dimension is derived from quantum Monte-Carlo calculations for
a 1D homogeneous electron gas where the electrons interact via the
soft-Coulomb interaction in
Eq.~(\ref{eq:softC})\cite{HFCVMTR2010}. The correlation energy is
parame\-trized in terms of $r_s$ and the spin polarization
$\zeta=(N_\uparrow-N_\downarrow)/N$ in the form
\bea
\label{eq:ectotal}
\epsilon_c(r_s,\zeta)&=&\epsilon_c(r_s,\zeta=0)\\
&&+\zeta^2
\left[\epsilon_c(r_s,\zeta=1)-\epsilon_c(r_s,\zeta=0)\right]
\nonumber
\eea
with
\bea
\nonumber
\epsilon_c(r_s,\zeta=0,1)&=&\!-\frac{1}{2}\frac{r_s + E r_s^2}
{A+ B r_s +C r_s^2 + D r_s^3 }\\
&&\times\ln ( 1 + \alpha r_s + \beta r_s^m )
\label{eq:ec}
\eea
which proves to be very accurate in the parameterization for 1D
systems for different long-range interactions
\cite{CSS2006,SCSM2009,HFCVMTR2010}. Note, that the above energy is
given in Hartree units. To obtain a priori the exact high-density
result known from the random-phase approximation, i.e.
\bea
\epsilon_c(r_s \rightarrow 0,\zeta=0) &=& -\frac{4}{\pi^4a^2}~r_s^2,\\
\epsilon_c(r_s \rightarrow 0,\zeta=1) &=& -\frac{1}{2\pi^4a^2}~r_s^2,
\eea
to leading order in  $r_s$, we fix the ratio $\alpha/A$ to be
equal to $8/(\pi^4 a^2)$ and $1/(\pi^4 a^2)$ for $\zeta=0$ and
$\zeta=1$, respectively. In both cases $m$ is limited to values
larger than 1. As a result, the number of independent parameters
in  Eq.\ (\ref{eq:ec}) is reduced to 7. In addition, for
$a=1$ the denominator can be simplified by setting $B=0$. However,
for smaller values of the softening parameter the linear term in
the denominator is important for achieving agreement with the
quantum Monte-Carlo results. The optimal values of the parameters
are given in Tab.~\ref{tab:lda}.  For more details on the 1D QMC
methodology and the parameterization procedure we refer to
Refs.~\cite{CSS2006,SCSM2009}.

\begin{table}
\setlength{\tabcolsep}{5mm}
\renewcommand\arraystretch{1}
\begin{center}
\begin{tabular}{|l|c|c|}\hline
& $\zeta=0$ & $\zeta=1$ \\ \hline \hline
A & 18.40(29)    & 5.24(79)    \\ \hline
B & 0.0          & 0.0         \\ \hline
C & 7.501(39)    & 1.568(230)  \\ \hline
D & 0.10185(5)   & 0.1286(150) \\ \hline
E & 0.012827(10) & 0.00320(74) \\ \hline
$\alpha$ & 1.511(24) & 0.0538(82) \\ \hline
$\beta$  & 0.258(6)  &  $1.56(1.31)\cdot 10^{-5}$ \\ \hline
m        & 4.424(25) &  2.958(99) \\ \hline\hline
$\Delta$ & $6.7\cdot 10^{-5}$ & $3.3\cdot 10^{-5}$ \\ \hline
\end{tabular}
\end{center}
\caption{\label{tab:lda}
Parameterization of the correlation energy of the 1D homogeneous
electron gas for a softening parameter of $a=1$, spin unpolarized
($\zeta=0$) and fully polarized ($\zeta=1$) cases are given. The
error in the last digits is given in parenthesis, while the average
error, $\Delta$, (in Hartree) in the full density range is given in the last row.}
\end{table}

We have implemented the 1D LDA for $a=1$ in both unpolarized and polarized
versions in the {\tt octopus} program \cite{MCBR2003,CAORALMGR2006}.

\begin{figure}
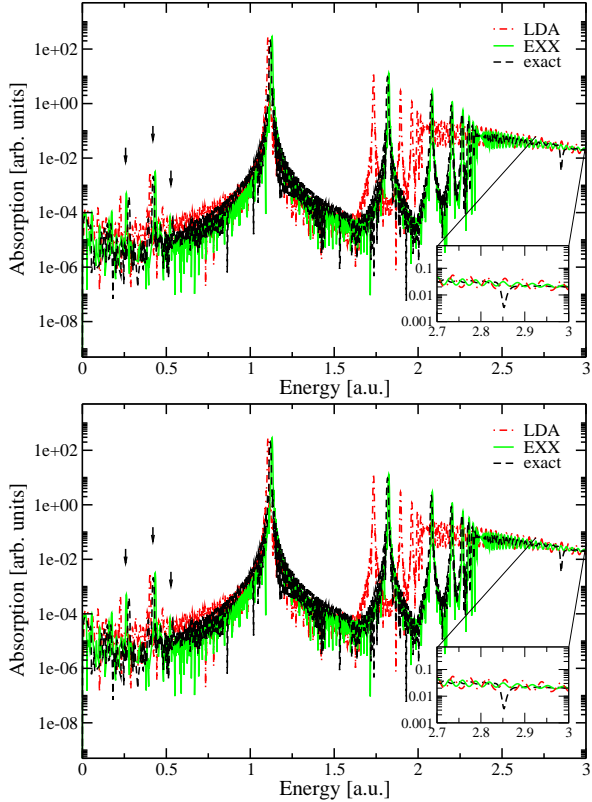

\includegraphics[width=0.47\textwidth, clip]{Be_lr}

\includegraphics[width=0.47\textwidth, clip]{Be_nlr}
\caption{
\label{fig:be2+} Linear (top) and non-linear (bottom) spectra of
Be$^{2+}$, calculated from the Fourier transformation of the dipole
moment using a polynomial damping function, comparing the exact and
the 1D LDA calculation. The inset in the bottom figure shows a zoom
into the region from $2.7$ to $3.0$ Ha.}
\end{figure}

Fig.~\ref{fig:be2+} shows the linear and non-linear absorption
spectra of a 1D Be$^{2+}$ system, i.e. with an external potential of
\be\label{eq:vextbe}
v_{\rm ext}^{\mathrm{Be}}(x)=\frac{-4}{\sqrt{x^2+a^2}}
\ee
containing two electrons. We use the LDA as an adiabatic approximation
to the exact time-dependent exchange-correlation potential. The
spectrum is calculated in linear response to a spatially constant
perturbation at $t=0$, i.e. we apply an additional external electric
field ${\cal E}$ in dipole approximation
\be\label{eq:kick}
v_{\rm ext}^{\mathrm{kick}}(x,t)=x{\cal E}_0\delta(t)
\ee
which gives an initial momentum to the electrons. The time
propagations were  performed in a box ranging from -150 to
150~bohr with absorbing boundary  conditions \cite{MCBR2003} and a
grid spacing of 0.2~bohr for a total propagation time of
$10^3$~a.u. In the linear regime a kick of ${\cal E}_0=10^{-4}$~Ha/bohr
was employed which was then increased to
$0.01$~Ha/bohr to obtain the non-linear response. The values of the
excitation energies can be found in Tab.~\ref{tab:be2+}. In linear
response, we see five peaks in the LDA spectrum which compare well
with the first five excitations in the exact case. As expected, the
agreement is better for lower lying excitations and gets worse the
closer we get to the continuum. As a guide for the eye we included the
KS HOMO energy of the LDA calculation, $\epsilon_{\mathrm{HOMO}}=2.06$
Ha and the exact ionization potential of 2.41 Ha. The onset of the
continuum itself appears at too low energies in the LDA calculation
missing two more clearly visible peaks in the exact spectrum. In other
words, the LDA fails to reproduce the proper Rydberg series, a
behavior well known from 3D calculations. For comparison we also
included the results from an EXX calculation (which for two electrons
is adiabatic and equal to Hartree-Fock) which shows a slightly better
agreement than LDA for the first three excitations but, more
importantly, reproduces the Rydberg series due to the correct
asymptotic behavior of the corresponding exchange potential. The
quality of the EXX results also implies that correlation is of
secondary importance in the system for $a=1$. The non-linear spectrum
shows the same excitations as the linear spectrum and three additional
peaks for the exact and the EXX calculation and two additional peaks
in the LDA spectrum. Their energies are also listed in
Tab.~\ref{tab:be2+}. Due to the spatial symmetry of the system all
even order responses are zero and the first non-vanishing higher-order
response is of third order. The frequency $\Omega_1=0.28$~Ha
corresponds to an excitation from the second to the third excited
state, where the transition from the ground to the second excited
state is dipole forbidden and, hence, can only be reached in a
two-photon process. The other two frequencies, $\Omega_2=0.42$~Ha and
$\Omega_3=0.54$~Ha, correspond to the transitions from first to second
and second to fifth excited state, respectively. Again, both the EXX
and the LDA calculations yield a good description of the low lying
excitations, only the third peak cannot be resolved in the LDA
spectrum.

\begin{table*}
\setlength{\tabcolsep}{2.2mm}
\renewcommand\arraystretch{1.15}
\begin{center}
\begin{tabular}{|l|ccccccc||ccc|}\hline
& $\omega_1$ & $\omega_2$ & $\omega_3$ & $\omega_4$ & $\omega_5$ & $\omega_6$ & $\omega_7$ & $\Omega_1$ & $\Omega_2$ & $\Omega_3$\\ \hline
LDA & 1.10 & 1.74 & 1.90 & 1.96 & 2.00 & - & - & 0.22 & 0.40 & -\\ \hline
EXX & 1.13 &  1.82 & 2.08 & 2.20 & 2.27 & 2.30 & 2.32 & 0.26 & 0.43 & 0.52\\ \hline
exact & 1.12 & 1.81 & 2.08 & 2.19 & 2.26 & 2.29 & 2.32 & 0.28 & 0.42 & 0.54\\
\hline
\end{tabular}
\end{center}
\caption{
\label{tab:be2+} Excitation energies from linear and non-linear
response of the 1D Be$^{2+}$ atom corresponding to the spectra in
Fig.~\ref{fig:be2+}. Excitations from linear response are denoted
as $\omega$ while those from the non-linear spectrum are denoted
with $\Omega$. All numbers are given in Hartree.}
\end{table*}

One feature of the exact spectrum that is missing from both the LDA
and the EXX spectra is the small dip at 2.8~Ha, see inset in
Fig.~\ref{fig:be2+}. It results from a Fano resonance
\cite{F1961,FC1968}, i.e.\ the decay of an excited state into
continuum states. It is missing from both approximate spectra due to
the double-excitation character of the involved excited state. Double
excitations in the linear regime can only be described in TDDFT if a
frequency-dependent xc kernel is employed \cite{MZCB2004}. Any
adiabatic approximation, however, leads to a frequency independent
kernel. Hence, double excitations, as well as any resulting features,
are missing from both the ALDA and the AEXX linear response
spectra. Apart from the well-known shortcomings of not including
double-excitations and not giving the correct Rydberg series, the 1D
ALDA reproduces both the linear and the non-linear exact spectra quite
well.

\begin{figure}
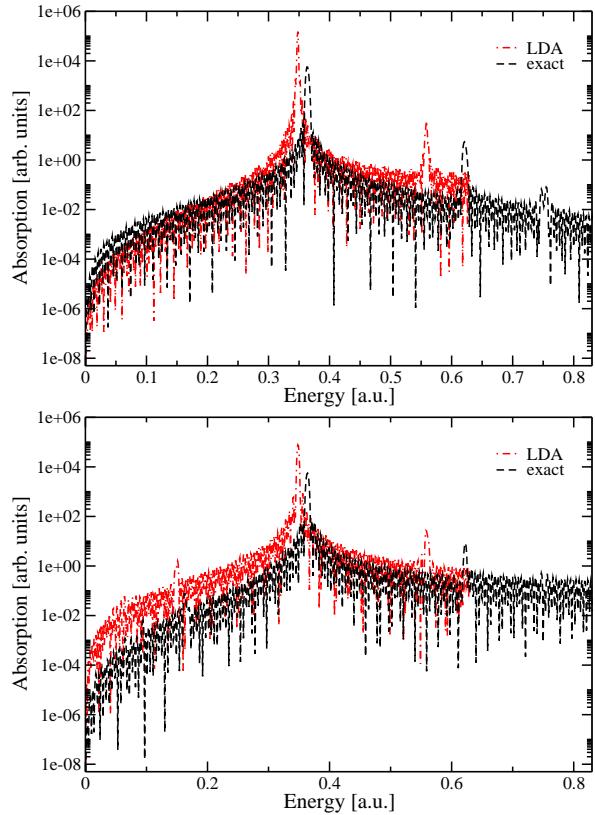

\includegraphics[width=0.47\textwidth, clip]{Beplus_lr}

\includegraphics[width=0.47\textwidth, clip]{Beplus_nlr}
\caption{\label{fig:be+}
Discrete part of the linear (top) and non-linear (bottom) spectra of
Be$^+$, calculated from the Fourier transformation of the dipole
moment using an exponential damping function, comparing the exact and
the 1D LDA calculation. The exact ionization potential is at 0.83~Ha
and the LDA HOMO at 0.63~Ha.}
\end{figure}

Fig.\ \ref{fig:be+} shows the discrete part of the linear and
non-linear spectra for the Be$^+$ system, i.e.\ the external potential
is given by Eq.\ (\ref{eq:vextbe}) and the system contains three
electrons. The ionization potential for the exact calculation is
0.83~Ha which is again underestimated by the LDA HOMO energy of
0.62~Ha. For this system the projection onto the Young diagrams
becomes important with the lowest energy spatial solution being
symmetric under exchange of any two variables.  Therefore, it is not a
valid solution for three fermions and, hence, discarded. The second
lowest energy is doubly degenerate with the eigenstates corresponding
to diagrams \ref{fig:young}c and \ref{fig:young}d. One of these states
is then propagated with a kick strength of ${\cal E}_0=10^{-4}$ for
the linear spectra and ${\cal E}_0=0.1$ for the non-linear spectra.
The exact linear spectrum shows two transitions at 0.36~Ha and
0.62~Ha.  Again, we observe that LDA underestimates these excitation
energies giving 0.34~Ha and 0.55~Ha, respectively.  The non-linear
spectrum contains two more peaks in the exact spectrum at 0.09~Ha and
0.16~Ha which are, however, difficult to resolve. In the LDA spectrum
only the peak at 0.16~Ha can be resolved. From the exact calculation
of excited states we know that there should be several more
transitions at very small frequencies which are accessible in
non-linear response.  Those are, however, very close to each other
and, hence, more difficult to resolve. Attempts to improve the spectra
in the small frequency region are currently in progress.

\section{Double excitations}\label{sec:double}

\begin{figure}
\begin{center}
\includegraphics[width=0.47\textwidth]{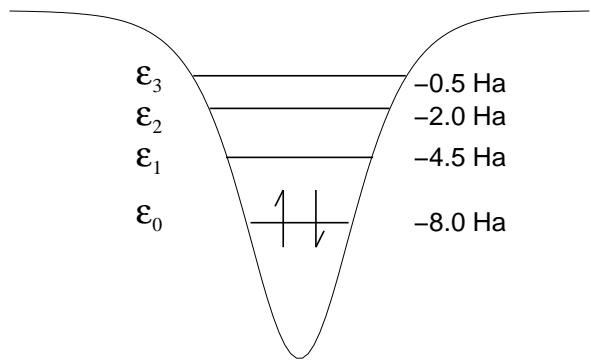}
\caption{\label{fig:extcosh}
External $\cosh$-potential and single particle eigenvalues. In the
non-interacting two-particle ground state both electrons occupy the
lowest energy level in a singlet configuration.}
\end{center}
\end{figure}

In order to investigate double excitations in the linear-response
spectrum we employ the following external potential
\be\label{eq:extcosh}
v_{\rm ext}^{\mathrm{cosh}}(x)=-\frac{v_0}{\cosh^2(kx)},
\ee
for which the one-particle problem can be solved analytically \cite{LL1977} and
the resulting eigenvalues are given as
\be
\epsilon_j=-\frac{k^2}{8}\left(\sqrt{1+\frac{8v_0}{k^2}}-1-2j\right)^2
\ee
for $j=0,1...$. Here, the term in parenthesis needs to be positive
which restricts the number of bound states of the system. In other
words, by choosing the two parameters $v_0$ and $k$ appropriately,
one can  create systems with any number of bound states. For our
calculations we choose $k=1$ and $v_0=10$ which leads to the four bound
single-electron states shown in Fig.~\ref{fig:extcosh}.
\begin{table*}
\begin{center}
\begin{tabular}{|r||c|c||c|c||c|}\hline
& \multicolumn{2}{|c||}{Symmetric well} & \multicolumn{2}{|c||}{Asymmetric well} & \\
State $j$ & non-interact. & interacting & non-interact. & interacting & Spin\\ \hline\hline
0 & -16.00 & -15.10 & -16.92 & -16.02 & singlet\\
1 & -12.50 & -11.75 & -13.21 & -12.45 & singlet\\
2 & -12.50 & -11.62 & -13.21 & -12.32 & triplet\\
3 & -10.00 & -9.31  & -10.52 & -9.83  & singlet\\
4 & -10.00 & -9.30  & -10.52 & -9.81  & triplet\\
5 & -9.00  & -8.22  & -9.48  & -8.70  & singlet\\
6 & -8.50  & -7.98  & -8.90  & -8.38  & singlet\\
7 & -8.50  & -7.97  & -8.90  & -8.38  & triplet\\
8 & -8.00  & -7.90  & -8.45  & -8.36  & singlet\\
9 & -8.00  & -7.90  & -8.45  & -8.36  & triplet\\ \hline
\end{tabular}
\end{center}
\caption{\label{tab:coshen}
Energies (in Hartree) for two-particle states in the symmetric well
potential (\ref{eq:extcosh}) and the asymmetric well
Eq.~(\ref{eq:extcoshmod}) without interaction and with soft-Coulomb
interaction, Eq.~(\ref{eq:softC}). The 5th excited state corresponds
to a double excitation.}
\end{table*}

\begin{table*}
\begin{center}
\begin{tabular}{|c||c|c|c||c|c|}\hline
& \multicolumn{3}{|c||}{Symmetric well} &
\multicolumn{2}{|c|}{Asymmetric well}\\
Excitation & Symmetry & non-interact. & interact. & non-interact. & interact.\\ \hline\hline
$\Omega_{01}$ & odd  & -3.50 & -3.48 & -3.71 & -3.70 \\
$\Omega_{02}$ & even & -6.00 & -5.80 & -6.40 & -6.21 \\
$\Omega_{11}$ & even & -7.00 & -6.88 & -7.44 & -7.32 \\
$\Omega_{03}$ & odd  & -7.50 & -7.13 & -8.02 & -7.64 \\
$\Omega_{12}$ & odd  & -9.50 & $\sim$ -9.28  & -10.11 & $\sim$ -9.91\\ \hline
\end{tabular}
\end{center}
\caption{\label{tab:coshOmegas}
Excitation energies (in Hartree) for two particles in the symmetric
well potential (\ref{eq:extcosh}) and the asymmetric well
Eq.~(\ref{eq:extcoshmod}) without interaction and with soft-Coulomb
interaction, Eq.~(\ref{eq:softC}). $\Omega_{11}$ and $\Omega_{12}$ are
double excitations. We also state the symmetry of the two-particle
excited state for the symmetric well.}
\end{table*}

\begin{figure}
\begin{center}
\includegraphics[width=0.47\textwidth]{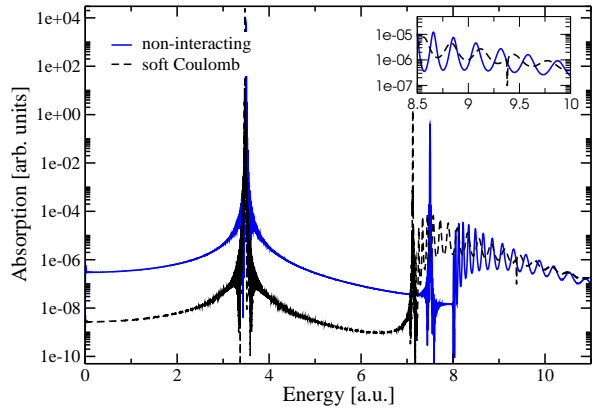}
\end{center}
\caption{\label{fig:linressymm}
Linear response spectrum for two electrons in a $\cosh$ potential
calculated from the Fourier transformation of the dipole moment using
a polynomial damping function. For interacting electrons we observe an
additional transition at $\approx 9.4$~Ha that corresponds to
$\Omega_{12}$ (see inset).}
\end{figure}

Putting two electrons into our system we calculate the total energies
for non-interacting electrons as well as for electrons interacting via
the soft-Coulomb interaction (\ref{eq:softC}). The results for the
first ten states are given in Tab.~\ref{tab:coshen}. As the energy
differences between the interacting and non-interacting cases are
small we can treat the many-body states as perturbed
independent-particle states.  This treatment is convenient because in
the independent-particle picture double excitations are well defined:
they describe transitions in which two electrons get excited with the
excitation energies given as the sum of two single-particle
excitations. The excitation energies for the different transitions are
shown in Tab.~\ref{tab:coshOmegas}. We note that the two-particle
eigenstates of the symmetric well (\ref{eq:extcosh}) can be chosen as
eigenstates of the parity operator and, hence, can be classified as
even and odd. For odd operators like the dipole operator the
transitions from the ground state (even) to even two-particle excited
states have zero oscillator strength. Nevertheless, these transitions
can be visible beyond linear response. In addition, starting from the
non-interacting ground state, doubly excited states have zero weight
in the density response function because the density operator is a
single particle operator. Thus, also the odd doubly excited
two-particle states have zero oscillator strength for non-interacting
particles. The 5th excited state of the non-interacting electrons can
clearly be identified as a double excitation. This transition
corresponds to both electrons getting promoted to the first excited
state $\varepsilon_1$. As the energy differences between the
interacting and non-interacting cases are small the 5th excited state
of the interacting system is of double-excitation character as
well. Unfortunately, however, this two-particle excited state is even
under parity and, hence, the transition is dipole forbidden. The first
double excitation which is dipole allowed is $\Omega_{12}$ which, in
the independent-particle picture, corresponds to one electron getting
promoted to the first excited state $\varepsilon_1$ and the other to
the second excited state $\varepsilon_2$. It has an excitation energy
of $9.50$~Ha in the non-interacting system. The first ionization
potential of the system, however, is $8.00$~Ha which implies that the
dipole allowed double excitation lies in the continuum part of the
spectrum. In addition, $\Omega_{12}$ has zero weight in linear
response for the non-interacting system because the transition matrix
element vanishes as the final state differs from the initial state in
two orbital occupations. For the interacting system, however, the
two-particle spatial singlet wave function is no longer given as a
product of the lowest energy single-particle orbital. A configuration
interaction (CI) expansion of this wave function also contains terms
which correspond to single excitations of the non-interacting
particles. As a result, the double excitation $\Omega_{12}$ becomes
accessible in linear response. This can be seen in
Fig.~\ref{fig:linressymm}, where we plot the absorption spectrum of
the two-electron system both for interacting and non-interacting
electrons. The spectrum was calculated in linear response to a
spatially constant perturbation at $t=0$, see Eq.~(\ref{eq:kick}). We
observe that the interacting spectrum shows a small dip at $\approx
9.4$~Ha which is close to the energy difference between the ground
state and the dipole-allowed double excitation described earlier (as
this excitation lies within the continuum its energy cannot be
computed directly but an estimate can be found from $\Omega_{01} +
\Omega_{02}$). We can clearly see that the transition lies within the
continuum, or due to the calculation being done in a finite box,
within the excitations to box states. Due to the nearby excitations to
the continuum the frequency $\Omega_{12}$ of the bound transition is
shifted slightly \cite{F1961}. This excitations appears as a dip
rather than a peak in the spectrum due to the absorbing boundary
conditions which were employed in the calculation \cite{FC1968}.

\begin{figure}
\begin{center}
\includegraphics[width=0.47\textwidth]{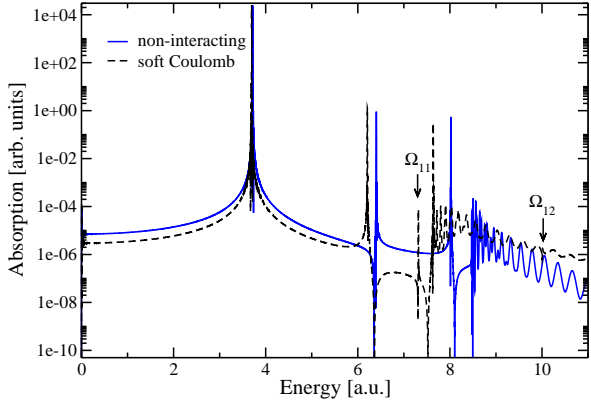}
\end{center}
\caption{\label{fig:linresasymm}
Linear response spectrum for two electrons in the modified $\cosh$
potential, Eq.~(\ref{eq:extcoshmod}), calculated from the Fourier
transformation of the dipole moment using a polynomial damping
function. Notice that for interacting electrons the double excitation
$\Omega_{11}$ appears at $\approx 7.3$ Ha and $\Omega_{12}$ shifts to
$\approx 10.0$ Ha.}
\end{figure}

\begin{figure}
\begin{center}
\includegraphics[width=0.47\textwidth]{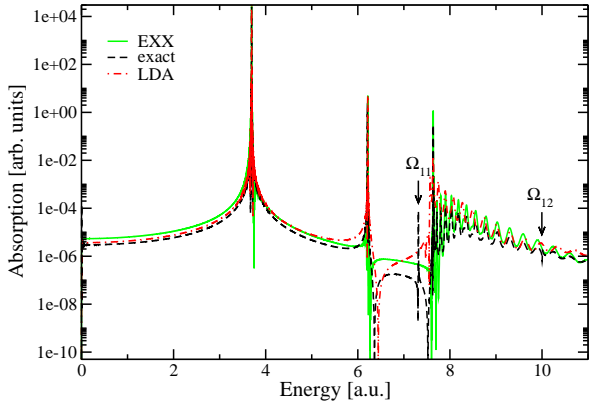}
\end{center}
\caption{\label{fig:linresasymmdft}
Linear response spectrum for two electrons in the modified $\cosh$
potential, Eq.~(\ref{eq:extcoshmod}), calculated from the Fourier
transformation of the dipole moment using a polynomial damping
function. For the DFT (EXX and LDA) spectra both double excitations
$\Omega_{11}$ and $\Omega_{12}$ are missing.}
\end{figure}

\begin{figure}
\begin{center}
\includegraphics[width=0.47\textwidth]{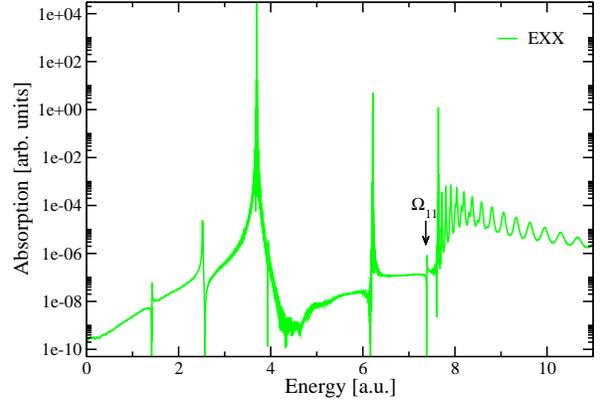}
\end{center}
\caption{\label{fig:nonlinresasymmdft}
Non-linear response spectrum for two electrons using EXX in the
modified $\cosh$ potential, Eq.~(\ref{eq:extcoshmod}), calculated from
the Fourier transformation of the dipole moment using a polynomial
damping function. The double excitation $\Omega_{11}$ appears at
$\approx 7.4$~Ha (see arrow).}
\end{figure}

In order to investigate the double excitations which are dipole
forbidden by symmetry, we break the spatial symmetry of the system by
modifying the external potential to
\be\label{eq:extcoshmod}
v_{\rm ext}^{\mathrm{mod}}(x)=-\frac{v_0(1+0.5x)}{\cosh^2(kx)}.
\ee
As the Hamiltonian for this case no longer commutes with the parity
operator, the eigenstates do not have a specific symmetry any
longer. Therefore, the previously dipole forbidden transitions now
have a finite oscillator strength. Also, the modification is small
enough to leave the ordering of the states intact, i.e. the fifth
excited state still has double-excitation character and the energies
are approximately those of the symmetric system (see
Tab.~\ref{tab:coshen} for details). As we can see in
Fig.~\ref{fig:linresasymm} this leads to an additional peak slightly
above 7.3~Ha which corresponds to the transition $\Omega_{11}$ from
the ground state to the fifth excited state.  In
Fig.~\ref{fig:linresasymmdft} we also include DFT linear response
spectra for the symmetry-broken potential Eq.~(\ref{eq:extcoshmod})
using EXX and the 1D LDA functional of section \ref{sec:lda} which are
used as adiabatic approximations to the time-dependent xc
potential. It is, therefore, not surprising that the double excitation
$\Omega_{11}$ is missing from the resulting spectra
\cite{MZCB2004,TK2009}.  However, as we can see in
Fig.~\ref{fig:nonlinresasymmdft}, beyond linear response the double
excitation $\Omega_{11}$ becomes visible even for the EXX functional,
which is adiabatic, in the symmetry-broken potential.

\section{Rabi oscillations}

Rabi oscillations can occur when a system is exposed to an external
laser field with frequency $\omega$, which is in resonance to a
transition in the system. The system oscillates between two states
with frequency $\Omega_0$ provided it can approximately be described
as a two-level system. This is the case if the frequency $\omega$ is
resonant to one specific transition in the system and the frequency of
the oscillation between the states is much smaller than the frequency
of the applied laser field. As will be discussed later, a small
detuning of the applied laser from the exact resonant frequency still
leads to Rabi oscillations, however, with increased frequency and
smaller amplitude.

We analyze Rabi oscillations for a 1D two-electron model (see section
\ref{sec:1D}) but note that the analysis can easily be extended to
three dimensions and, using the single-pole approximation also to any
number of electrons \cite{FHTR2011}. For ease of comparison we choose
the same model as in \cite{BauerRugg} with the external potential
\be
v_{\rm ext}^{\mathrm{Rabi}}(x,t)= -\frac{2}{\sqrt{x^2 + 1}}+x{\cal E}_0\sin(\omega t).
\label{eq:vextrabi}
\ee
Eq.~(\ref{eq:vextrabi}) describes a 1D Helium atom interacting with a
monochromatic laser field of frequency $\omega$.  We denote the
eigenstates and eigenvalues of the Hamiltonian (\ref{eq:1dham}) with
the external potential (\ref{eq:vextrabi}) as $\psi_k$ and
$\epsilon_k$, respectively. In order for Rabi's solution to be valid,
the system under study needs to be an effective two-level system
reducing the solution space to $\psi_0$ (ground state) and $\psi_1$
(dipole allowed excited state) with eigenenergies $\epsilon_0$ and
$\epsilon_1$. The system then has a resonance at
$\Delta=\epsilon_1-\epsilon_0$.  The two-level approximation is valid
if the two conditions
\begin{equation}
 \frac{\delta}{\Delta}<< 1, \quad  \Omega_0 << \omega
\label{eq:cond}
\end{equation}
are satisfied, where $\delta= \omega -\Delta$ describes the detuning
from the resonance and $\Omega_0=d_{10}{\cal E}_0$ is the Rabi
frequency for a resonant laser, with $d_{10}=\langle \psi_0 | \sum_j
\hat{x}_j | \psi_1 \rangle$ the dipole matrix element. In order to
satisfy the second condition we choose ${\cal E}_0=0.0125\omega$ in
Eq.~(\ref{eq:vextrabi}). For the eigenvalues we obtain
$\epsilon_0=-2.238$ Ha and $\epsilon_1=-1.705$ Ha which implies a
resonant frequency of $\Delta=\epsilon_1-\epsilon_0=0.534$~Ha. The
static dipole matrix element is $d_{10}=1.104$. The frequency
$\omega$ of the applied field has been chosen to be close to the
resonance, i.e. the detuning $\delta=\omega - \Delta$ is small.

For an effective two-level system the time-dependent two-electron wave
function $\psi(x_1, x_2, t)$ can be written as a linear combination of
ground and excited state, i.e.
\be \label{eq:2levellin}
\psi(x_1,x_2,t) = a_0(t) \psi_0(x_1,x_2)+ a_1(t) \psi_1(x_1,x_2)
\ee
with $|a_0(t)|^2=n_0 (t)$ and $|a_1(t)|^2=n_1(t)$ being the
time-dependent  level populations of the ground and excited
states. Normalization of the wave functions then implies $n_0(t) +
n_1(t) = 1$. As the amplitude ${\cal E}_0$ and  the frequency
$\omega$ of the applied field ${\cal E}(t)$ are chosen such that
the conditions (\ref{eq:cond}) are fulfilled the Hamiltonian
(\ref{eq:1dham}) can be projected onto a $2\times 2$ space. The
time-dependent Schr\"odinger equation $i \partial_t
|\psi(t)\rangle = \hat{H} |\psi(t)\rangle $ then reduces to a
$2\times 2$ matrix equation of the form
\be
i \partial_t  \left( {\begin{array}{c}
 a_0(t)\\
 a_1(t)  \\
 \end{array} } \right)=
\left( {\begin{array}{cc}
 \epsilon_0  &  d_{10}{\cal E}(t)  \\
d_{10}{\cal E}(t) & \epsilon_1  \\
 \end{array} } \right) \left( {\begin{array}{c}
 a_0(t)   \\
 a_1(t)
 \end{array} } \right)
\label{eq:proj1}
\ee
from which one can derive coupled differential equations for the level
population $n_1(t)$ and the dipole moment
$d(t)=\langle \psi(t)|\hat{x}_1+\hat{x}_2 |\psi(t) \rangle =
2d_{10} Re\left(a_0^*(t)a_1(t)\right)$. For the dipole moment we obtain
\begin{equation}
d(t)= 2d_{10}\sqrt{n_0(t)n_1(t)}\cos(\omega t + \theta(t)),
\label{eq:d_t_lin}
\end{equation}
where the Rabi frequency is included in the time dependence of the
amplitude of the dipole moment through the time dependence of $n_0(t)$
and $n_1(t)$, and $ \theta(t)$ is the phase difference of $a_0(t)$ and
$a_1(t)$.  Fulfillment of conditions (\ref{eq:cond}) allows for the
use of the rotating wave approximation (RWA) \cite{QMTD} which yields the
following differential equation for $n_1(t)$
\be
\partial^2_t n_1(t)= -\left(\delta^2 + \Omega_0^2\right) n_1(t) + \frac{1}{2}\Omega_0^2
\label{eq:n_1_t_t_lin}
\ee
with initial conditions $n_1(0)=0$ and $\dot{n}_1(0)=0$.
Eq.~(\ref{eq:n_1_t_t_lin}) describes a harmonic oscillator with a
restoring force which increases with increasing detuning $\delta$. As
a result, the frequency of the Rabi oscillations increases with
increased detuning, while the maximum population of the excited state
decreases as $n_1^{\mathrm{max}}=\Omega_0^2/(\Omega_0^2+\delta^2)$.

\begin{figure}
\begin{center}
\includegraphics[width=0.47\textwidth]{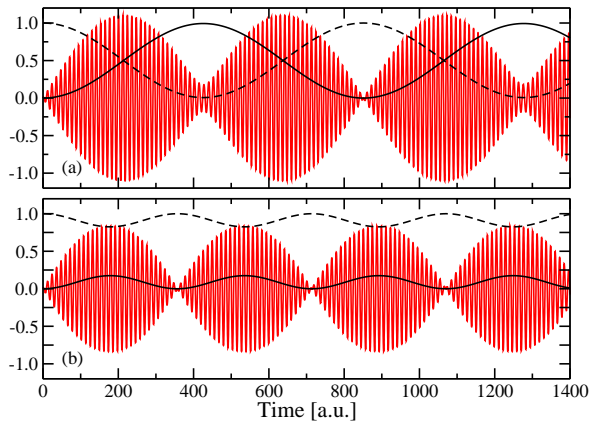}
\end{center}
\caption{\label{fig:dipolen1Lin}
Dipole moment (red) and level populations $n_1(t)$ (solid black line) and
$n_0(t)$ (dashed black line) from analytic solution of (\ref{eq:proj1})
using ${\cal E}_0= 0.0125 \omega$ for detuning $\delta=0.08\:
\Omega_0$ ($0.0006$ Ha) (a) and $\delta=2.2\: \Omega_0$ ($0.016$ Ha)
(b).}
\end{figure}

In Fig.~\ref{fig:dipolen1Lin} the time-dependent dipole moment $d(t)$ and the
level populations $n_0(t)$ and $n_1(t)$ for $\delta= 0.08\: \Omega_0$ and
$\delta=2.2\: \Omega_0$ are shown. The effect of the detuning manifests itself
in an incomplete population of the excited state and a consequent decrease in
the amplitude of the envelope of the dipole moment that is proportional to
$\sqrt{n_0 n_1}$. For small detuning the minima and the maxima of $n_1$ coincide
with minima of the envelope, but for larger detuning the dipole moment only goes
to zero for the minima of $n_1$. In Fig.\\ref{fig:dipolen1Lin}a the detuning is
small but non-zero, hence, the neck at the minima of $n_0$.  The neck grows with
increasing $\delta$ and evolves into a maximum for 
Fig.~\ref{fig:dipolen1Lin}b.  Thus, the first minimum in
Fig.~\ref{fig:dipolen1Lin}b corresponds to one complete cycle and can be
identified with the second minimum in Fig.~\ref{fig:dipolen1Lin}a. We note that
looking only at the dipole moment is insufficient to discern between resonant
and detuned Rabi oscillations, only studying the level populations gives the
complete picture.

For the system specified above a comparison between the analytic
solution of Eqs.~(\ref{eq:d_t_lin}), (\ref{eq:n_1_t_t_lin}) and the
results of the time-propagation with the {\tt octopus} code
\cite{MCBR2003,CAORALMGR2006} shows a perfect agreement, which
confirms that the conditions (\ref{eq:cond}) are fulfilled for the
chosen values of ${\cal E}_0$ and $\omega$.


The KS Hamiltonian corresponding to (\ref{eq:vextrabi}) is given as
\begin{equation}
H_s= H_s^0 + \sum_{j=1}^N\left(v_{\rm hxc}^{\rm dyn}(x_j, t) + x_j {\cal E}_0\sin(\omega
t)\right),
\label{eq:NonLinH}
\end{equation}
where the static KS Hamiltonian reads
\be
H_s^0= \sum_{j=1}^N -\frac{\nabla_j^2}{2}+ v_{\rm ext}(x_j) +
v_{\rm hxc}[\rho_0](x_j).
\ee
We denote the eigenfunctions of $H_s^0$ with $\phi_k(x)$ and their
eigenvalues as $\epsilon_k^{s}$. As we are using a two-electron
system, a single orbital is doubly occupied in the KS system. The time
evolution of this orbital follows from the KS equation
\be
i \partial_t \phi(x, t) = H_s \phi(x, t)
\ee
with the initial condition $\phi(x, t=0)=\phi_0(x)$. This equation is
non-linear due to the dependence of the Hartree-exchange-correlation
potential $v_{\rm hxc}$ on the density, 
$\rho(x,t)= 2 |\phi(x,t)|^2$. The time-dependent dipole moment $d(t)$
is an explicit functional of the density,
i.e. $d(t)=\int x \rho(x,t)\,dx$.  The exact KS system reproduces the
exact many-body density $\rho(x,t)$ and, hence, the exact dipole
moment $d(t)$. However, this need not be true for an approximate
functional.  Especially, using adiabatic approximations has been shown
to have a dramatic effect on the calculated density during Rabi
oscillations \cite{BauerRugg}.


\begin{figure}
\begin{center}
\includegraphics[width=0.47\textwidth]{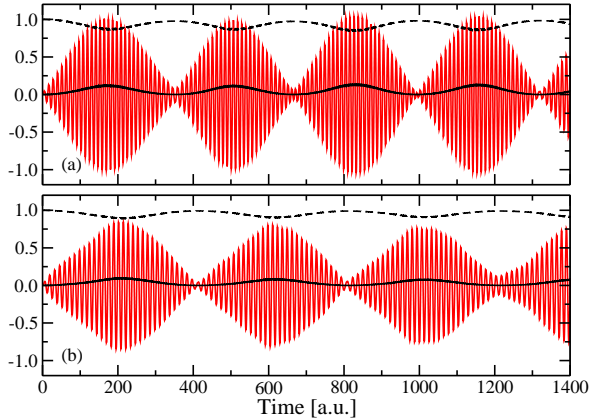}
\end{center}
\caption{\label{fig:ExxLDA} Dipole moment (red) and level populations
$n_1^s$ (solid black line) and $n_0^s$ (dashed black line) for EXX (a) and ALDA (b).}
\end{figure}

Propagating with the EXX and ALDA (see section \ref{sec:lda}) results in the
dipole moments shown in Fig.~\ref{fig:ExxLDA}.  The resonant frequencies are
calculated from linear response which yields $\omega^{ALDA}=0.476$ Ha and
$\omega^{EXX}= 0.549$ Ha. We then apply a laser field in analogy to the exact
calculation with an amplitude of ${\cal E}_0=0.0125\omega$ using the resonant
frequency for each case. We observe in the level populations that both EXX and
ALDA show the characteristic signatures of detuned Rabi oscillations despite the
fact that the applied laser is in resonance with the system.


To clarify whether Rabi oscillations are well described in the
context of adiabatic TDDFT we study the Hamiltonian
(\ref{eq:NonLinH}) in more detail. The Hartree and xc potentials,
$v_{\rm hxc}[\rho]$, render the KS differential equation non-linear.
More specifically, for an adiabatic approximation the potential at
time $t$ is a functional of the density at this time, i.e.
$v_{\rm hxc}(t)=v_{\rm hxc}[\rho(t)]$. In the following, we show that
$v_{\rm hxc}(t)$ introduces a detuning that drives the system out of
resonance. We again rely on the conditions (\ref{eq:cond}), i.e.\
describe the KS system as an effective two-level system.
Therefore, the time-dependent orbital $\phi(x, t)$ is given as a
linear combination of the ground-state KS orbital $\phi_0$ and the
first excited state orbital $\phi_1$
\be \label{eq:2level}
\phi(x,t) = a_0^s(t) \phi_0(x) + a_1^s(t) \phi_1(x).
\ee
Projecting the KS Hamiltonian (\ref{eq:NonLinH}) onto the two-level KS
space (\ref{eq:2level}) yields the $2\times 2$ matrix
\be
\left( {\begin{array}{cc}
 \epsilon_0^s + \epsilon_0^{xc}(t) &  d_{10}^s{\cal E}(t) + {\cal F}_{xc}(t) \\
d_{10}^s{\cal E}(t) + {\cal F}_{xc}^*(t) & \epsilon_1^s +
\epsilon_1^{xc}(t) \\
 \end{array} } \right)
\label{eq:proj2}
\ee
with $d_{10}^s=\langle \phi_1|\hat{x} |\phi_0 \rangle$,
$\epsilon_j^{xc}(t)=\langle \phi_j|\hat{v}_{hxc}^{\rm dyn}(t)|\phi_j
\rangle$, and ${\cal F}_{xc}(t)=\langle \phi_0| \hat{v}_{hxc}^{\rm
  dyn}(t)|\phi_1 \rangle$. This matrix enters Eq.~(\ref{eq:proj1}) to
determine the coefficients $a_0^s(t)$ and $a_1^s(t)$. Compared to
Eq.~(\ref{eq:proj1}) we notice that each entry contains an additional
term depending on $v_{\rm hxc}^{\rm dyn}$, i.e.\ both the electric
field and the KS eigenvalues are modified. In order to investigate the
consequences of the additional terms we use the EXX functional for
which relatively simple analytic expressions for the additional matrix
elements can be derived.

For the two-electron case investigated here, the
Hartree-exchange-correlation potential $v_{\rm hxc}^{EXX}(x,t)$ is
equal to half the Hartree potential and, hence, given as
\begin{equation}
v_{\rm hxc}^{EXX}(x,t)= \frac{1}{2}\int \frac{\rho_0(x') + \delta
\rho(x',t)}{\sqrt{(x-x')^2+a^2}} dx',
\label{eq:VHXCEXX}
\end{equation}
where we split the total density $\rho(x, t)$ into a
time-independent contribution $\rho_0(x)=2|\phi_0(x)|^2$ and a
rest $\delta \rho(x, t)$ which can be calculated from
Eq.~(\ref{eq:2level}) as
\bea\label{eq:deltarho}
\delta \rho(x,t) &=& 2|a_1^s|^2 ( |\phi_1(x)|^2 - |\phi_0(x)|^2)\\
&&\nonumber
+ 4 d_{10}^s Re(a_0^s(t)*a_1^s(t)) \phi_1(x)\phi_0(x).
\eea
The part of Eq.~(\ref{eq:VHXCEXX}) containing $\rho_0$ determines
$v_{\rm hxc}[\rho_0]$ while $\delta\rho$ results in the additional
$v_{\rm hxc}^{\rm dyn}$. Since $\phi_0$ and $\phi_1$ usually have
opposite spatial symmetry (if the Hamiltonian is symmetric) the first
term in Eq.~(\ref{eq:deltarho}) is symmetric while the second is
antisymmetric. This results in the first term only contributing to the
diagonal elements of the matrix (\ref{eq:proj2}) while the second term
only contributes to the off-diagonal elements. Defining the level
populations in the KS system as $n_j^{s}(t)= |a_j^s(t)|^2$ allows us
to rewrite the contributions to the diagonal terms as
\begin{equation}
\epsilon_j^{xc}(t)=\lambda_j n_1^{s}(t),
\end{equation}
where, for the EXX approximation, the coefficient $\lambda_j$
reads
\begin{equation}
\lambda_j = \int\!\!\int
\frac{(|\phi_1(x')|^2 -|\phi_0(x')|^2)|\phi_j(x)|^2}
{\sqrt{(x-x')^2+a^2}} dx dx'.
\end{equation}
For the off-diagonal contribution we recall that $d^{s}(t)=
2d_{10}^{s}Re(a_0^s(t)^*a_1^s(t))$, as in the exact case, and rewrite
the contribution of $v_{\rm hxc}^{\rm dyn}(t)$ to the off-diagonal
terms as a coefficient $g$ multiplied by the time-dependent dipole
moment
\begin{equation}
{\cal F}_{xc}(t)= g \frac{d^{s}(t)}{d_{10}^s}.
\end{equation}
For the two-electron system in the EXX approximation $g$ is given as
\begin{equation}
g= \int\!\!\int
\frac{\phi_1(x')\phi_0(x') \phi_0(x)\phi_1(x)}{\sqrt{(x-x')^2+a^2}} dx dx'.
\label{eq:g}
\end{equation}
The coefficient $g$ also enters the calculation of the resonant
frequencies in linear response. Within the single-pole approximation
the resonant frequency is given as $\omega_0^{s}=\sqrt{\Delta_s
  (\Delta_s + 2g)}$ which for the EXX functional yields $\omega^{EXX}=
0.532$~Ha.  The deviation from the frequency calculated from time
propagation of the Hamiltonian (\ref{eq:NonLinH}) in {\tt octopus} is
of the order of $3\%$, coinciding with the deviation of our system
from a true two-level system which we estimate from
$\left(1-(n_0^s(t)+n_1^s(t))\right)$. Using, as in the exact case, the
RWA we obtain to leading order in $\lambda/\omega_0$ and $g/\omega_0$
the following equation of motion for the level population $n_1(t)$
\be
\partial^2_t n_1^s(t) =  - \left(\frac{\gamma^2}{2}n_1^s(t)^2 +
\Omega_s^2\right) n_1^s(t) +\frac{1}{2}\Omega_s^2
\label{eq:n1_t_t}
\ee
with $\Omega_{s}=d_{10}^{s}{\cal E}_0$ and $\gamma=\lambda -
2g$. Neglecting the higher order terms in $\lambda/\omega_0$ and
$g/\omega_0$ introduces an error of about 10\% in favor of keeping the
equation simple while still containing the important physical
effects. Unlike Eq.~(\ref{eq:n_1_t_t_lin}) which represents a harmonic
oscillator, Eq.~(\ref{eq:n1_t_t}) corresponds to an anharmonic quartic
oscillator and its solution is given in terms of Jacobi elliptic
functions \cite{AQO}. Equivalently, Eq.~(\ref{eq:n1_t_t}) can be
integrated numerically. Even though the oscillator is no longer
harmonic the detuning still results in an increase of the restoring
force. In other words, the adiabatic approximation introduces a
time-dependent detuning proportional to $\gamma n_1^s(t)$.

\begin{figure}
\begin{center}
\includegraphics[width=0.47\textwidth]{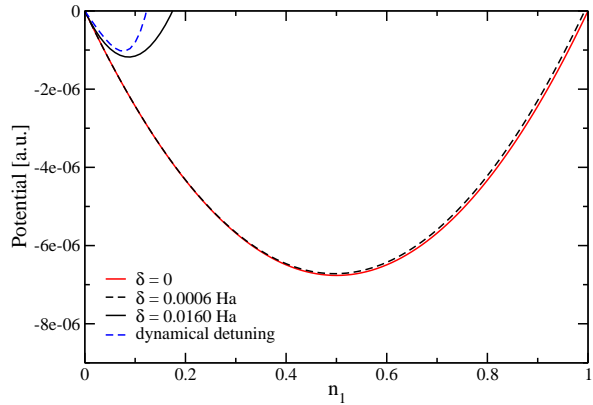}
\end{center}
\caption{\label{fig:rabipot}
Potentials corresponding to the differential equations
(\ref{eq:n_1_t_t_lin}) and (\ref{eq:n1_t_t}).  The dynamical detuning
leads to a quartic potential which has a similar effect to the
potential as a large detuning.}
\end{figure}

Using the same 1D model system (\ref{eq:vextrabi}) as before and
diagonalizing the Hamiltonian (\ref{eq:NonLinH}) gives for the bare KS
eigenvalues $\epsilon_0^{EXX}=-0.750$~Ha and
$\epsilon_1^{EXX}=-0.257$~Ha which yields $\Delta_s=0.494$~Ha.
For the various matrix elements we obtain $d_{10}^{s}= 0.897$,
$g=0.071$, $\lambda= -0.125$, and $\gamma=-0.268$. As a result
the detuning is of the order of 5\% of the resonant frequency,
i.e.\ quite large compared to the detuning which is necessary to
destroy the resonant Rabi behavior (see Fig.~\ref{fig:dipolen1Lin})
which explains the results we see for the dipole moment and level
populations (see Fig.~\ref{fig:ExxLDA}a). In Fig.~\ref{fig:rabipot} we
plot the potential corresponding to the restoring forces in the
differential equations (\ref{eq:n_1_t_t_lin}) and
(\ref{eq:n1_t_t}). As we can see, the dynamical detuning in the EXX
calculation has a similar effect on the squeezing of the potential as
the large detuning in the linear Rabi oscillations.

The behavior using ALDA is very similar to the one for the EXX
approximation (see Fig.~\ref{fig:ExxLDA}b). The analysis, however, is
more involved due to the functional not being linear in the
density. We can conclude that any adiabatic functional, even the exact
adiabatic one \cite{TGK2008}, will lead to a detuning in the
description of Rabi oscillations due to the lack of memory and the
fact that the exact density changes dramatically during the
transition. For the exact functional the detuning effect is
compensated by a dependence on the density at previous times.


\section{Reduced density-matrix functional theory}

\begin{figure*}[t]
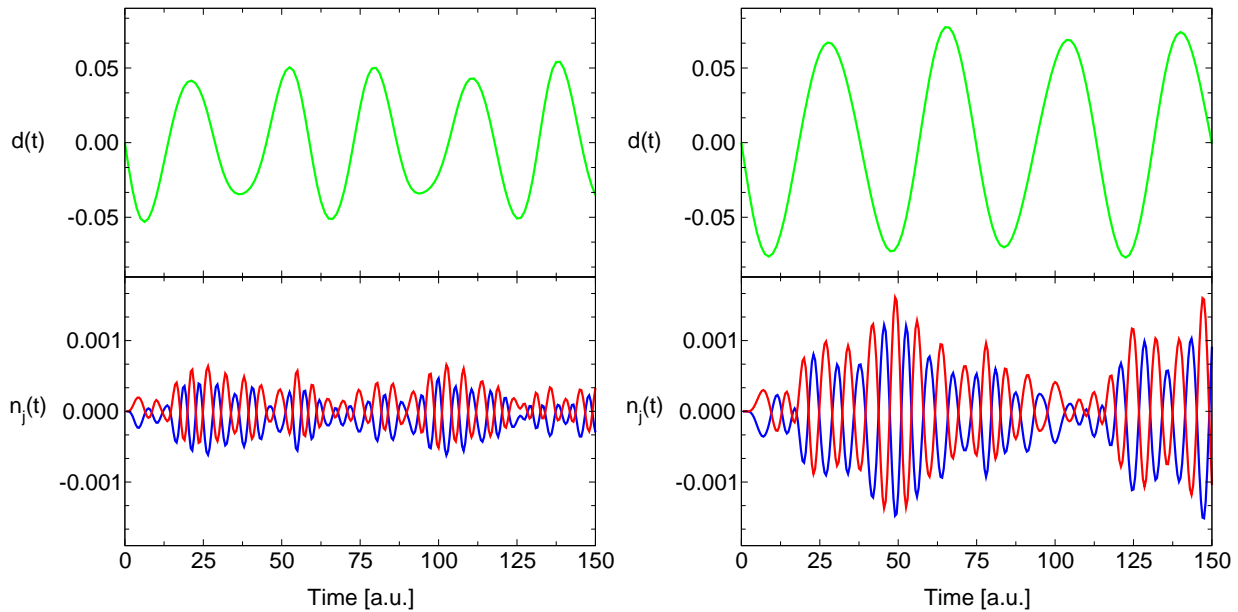

\begin{center}
\includegraphics[width=0.49\textwidth,clip]{occupations-he}
\includegraphics[width=0.49\textwidth,clip]{occupations-h2}
\end{center}
\caption{
  Changes in the dipole moment (green) and the first two natural occupation
  numbers (red and blue) for the helium atom (left) and the hydrogen
  molecule (right) in linear response to a $\delta$-kick of strength
  ${\cal E}_0=0.05$~Ha/bohr.
  \label{fig:h2kick}
  }
\end{figure*}

The use of adiabatic approximations in DFT is known to miss certain
aspects of the interacting system, one of which being the double
excitations discussed in Section \ref{sec:double}. Another example
are charge-transfer excitations which, while appearing in linear
response, are not described correctly \cite{DH2004}. Recently, it has
been suggested that using reduced density-matrix functional theory
(RDMFT) these problems can be addressed, even within adiabatic
approximations \cite{GPGB2009}.

RDMFT uses the one-body density matrix (1RDM)
\be
\gamma_1(\v r, \v r')=\!N\!\!\int\!\!\!\!\int\! d^3r_2...d^3r_N
\Psi(\v r, \v r_2...\v r_N)\Psi^*(\v r', \v r_2...\v r_N)
\ee
as its basic variable. Approximations within this theory are
usually stated in terms of the eigenfunctions $\varphi_j(\v r)$ and
eigenvalues $n_j$ of $\gamma_1(\v r, \v r')$ which  satisfy
\be
\int d^3r' \gamma_1(\v r, \v r')\varphi_j(\v r')=n_j\varphi_j(\v r)
\ee
and are called natural orbitals and occupation numbers,
respectively. The {\em exact} kinetic energy of a system can be
written explicitly in terms of the 1RDM which presents a major
advantage compared to standard density functional theory.

Since in general the interaction energy is known as an exact
functional of the diagonal $\gamma_2(\v r,\v r'; \v r, \v r')$
of the two-body density matrix (2RDM)
\begin{align}
\gamma_2(\v r_1,\v r_2; \v r'_1, \v r'_2)=\!\frac{N(N-1)}{2}\!\!\int\!\!\!\!\int\! d^3r_3...d^3r_N \\
\Psi(\v r_1, \v r_2, \v r_3 ...\v r_N)\Psi^*(\v r'_1, \v r'_2, \v r_3 ...\v r_N), \nonumber
\end{align}
RDMFT can be viewed as a way to express the diagonal of the 2RDM as a
functional of the 1RDM. Using this functional, the interaction energy
is written typically as a sum of the Hartree energy and the
exchange-correlation (xc) energy, where only the correlation energy
needs to be approximated. Approximations are usually stated by writing
either the diagonal of the 2RDM or the xc energy as a functional of
the natural orbitals and occupation numbers. For most currently
employed functionals the xc energy takes the form
\cite{M1984,GU1998,BB2002,GPB2005,P2006,SDLG2008}
\bea\label{eq:excrdm}
E_{xc}[\{\varphi_j\},\{n_j\}]&=&-\frac{1}{2}\sum_{j,k=1}^\infty f(n_j,n_k)\\
&&\nonumber\hspace*{-1.2cm}
\times\int\!\!\!\!\int \frac{\varphi_j(\v r)\varphi_j^*(\v r')
\varphi_k(\v r')\varphi_k^*(\v r)}{|\v r-\v r'|} d^3rd^3r',
\eea
i.e. it is given as an exchange integral modified by a function
depending on the occupation numbers. Generally, one can expand the 2RDM in the
basis of the natural orbitals
\be\label{eq:twordm}
\gamma_2(\v r_1,\v r_2; \v r'_1, \v r'_2)=\sum_{ijkl}
\gamma_{2,ijkl}\varphi_i(\v r_1)\varphi_j(\v r_2)\varphi_k^*(\v r'_1)\varphi_l^*(\v r'_2),
\ee
which gives rise to expansion coefficients
$\gamma_{2,ijkl}$. Restricting these coefficients to the form
\be\label{eq:twordmcoeff}
\gamma_{2,ijkl}=n_i n_j\delta_{ik}\delta_{jl}-f(n_i,n_k)\delta_{il}\delta_{jk},
\ee
the diagonal, $\gamma_2(\v r,\v r'; \v r, \v r')$, yields the Hartree
energy and the xc energy given in Eq.\ (\ref{eq:excrdm}).

Currently, an effort is made to extend the static theory in order to
describe time-dependent systems. The most straightforward extension is
again achieved by employing an adiabatic approximation
\cite{AG2010,PGGB2007}, i.e.\ the 2RDM at time $t$ is only treated as
a functional of the 1RDM at this point in time. Consequently, in
Eq.\ (\ref{eq:twordm}), $\gamma_2$ aquires a dependence on time $t$ as
do the occupation numbers and natural orbitals.  For the coefficients
of the 2RDM (in the basis of the natural orbitals at time $t$) such an
adiabatic approximation amounts to
\bea
\nonumber
\gamma_{2,ijkl}(t)\!\!\! &=& \!\!\!g_{ijkl}[\{n_j(t)\}]\delta_{ik}\delta_{jl}-h_{ijkl}[\{n_j(t)\}]\delta_{il}\delta_{jk}\\
&&+\lambda_{ijkl}[\{n_j(t)\}],
\label{eq:rdmftfunc}
\eea
where $g_{ijkl}(t)$ and $h_{ijkl}(t)$ are the time-dependent
coefficients for the Hartree and exchange-type integrals,
respectively. The time-dependent cumulant $\lambda_{ijkl}(t)$ contains
all contributions to the 2RDM which are not of Hartree or exchange
type. All three coefficients are approximated as functionals of the
occupation numbers. Comparing Eqs.\ (\ref{eq:twordmcoeff}) and
(\ref{eq:rdmftfunc}) we note, that an adiabatic extension of the
currently employed static functionals to the time domain leads to a
vanishing cumulant $\lambda$. By inserting Eq.\ (\ref{eq:rdmftfunc})
into the equation of motion for the natural occupation numbers
\cite{AG2010}, we find
\begin{equation}
\begin{split}
\label{eq:zerorhs}
i\, \partial_t n_k(t) &= \sum_{ijl}\lambda_{ijkl}(t)\langle ij|v_{\mathrm{int}}|kl\rangle (t)-c.c.,
\end{split}
\end{equation}
which directly illustrates that adiabatic approximations based on the
form of Eq.\ (\ref{eq:twordmcoeff}) cause a zero right-hand side and,
hence, lead to occupation numbers which are constant in time. While
one can imagine this to be a reasonable approximation in some
situations, it will generally not be the case. Including an explicit
cumulant in the approximation of $\gamma_{2,ijkl}(t)$ leads to
occupation numbers which can aquire a true time-dependence, even if
one chooses an adiabatic approximation for $\lambda_{ijkl}(t)$.

To assess the quality of the adiabatic approximation in RDMFT
quantitatively, we employ again our 1D model. For such model systems
with a small number of electrons one can extract the {\it exact} 1RDM
from the solution of the time-dependent Schr\"odinger equation of the
interacting system, which allows us to explicitly investigate for
which situations constant occupation numbers yield a reasonable
description. For this purpose, we employ two different two-electron
systems, a 1D helium atom and a 1D hydrogen molecule. The external
potentials are given by
\bea
v_{\rm ext}^{\mathrm{He}}(x)\!\!\!\!&=&\!\!\!\!-\frac{2}{\sqrt{x^2+a^2}}, \\
\nonumber
v_{\rm ext}^{\mathrm{H_2}}(x)\!\!\!\!&=&\!\!\!\!-\frac{1}{\sqrt{(x+d/2)^2+a^2}}
-\frac{1}{\sqrt{(x-d/2)^2+a^2}}\\
&&+\frac{1}{\sqrt{d^2+a^2}},
\eea
i.e. we are using a soft-Coulomb potential to describe the interaction
between the nuclei and the electrons and, for the hydrogen molecule,
also the interaction between the two nuclei.

In the following, we investigate two different sit\nolinebreak
uations.  In the first case we apply a kick, Eq.~(\ref{eq:kick}),
which provides an initial momentum to the system. We choose the
strength ${\cal E}_0=0.05$~Ha/bohr such that the evolution can be
described in linear response. As a second case, we investigate the
transition of the system, here the helium atom, from its ground state
to the first excited singlet state. To this end we use optimal control
theory \cite{RZ2000,SB2003,werschnik-2007} to find an optimized laser
pulse which induces a transition with a population of the excited
state of 98.59\% at the end of the pulse.

In Fig.~\ref{fig:h2kick}, we show the dipole moment and the change in
the first two natural occupation numbers, $\Delta
n_j(t)=n_j(t)-n_j(t=0)$, for both the helium atom and the hydrogen
molecule. The strength of the kick was chosen as the maximum possible
while staying within a linear response description. As we can see, the
occupation numbers show pronounced oscillations which, however, remain
small in amplitude compared to their ground state values while the
dipole moment shows the characteristic oscillations. Hence, in linear
response, a description with constant occupation numbers will be
appropriate.

\begin{figure}[t]
\begin{center}
\includegraphics[width=0.47\textwidth]{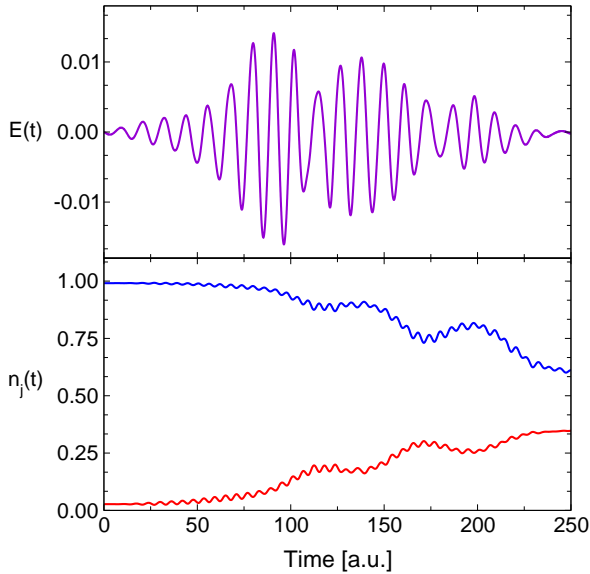}
\end{center}
\caption{
  The upper panel displays the laser amplitude of an optimized
  laser pulse that induces a transition from the ground state
  to the first excited singlet state of Helium. In the lower
  panel the two largest occupation numbers of the reduced
  one-body density matrix are shown as function of time.
  \label{fig:helium_occupations}
  }
\end{figure}

As an example where the occupation numbers change significantly, we
examine the transition of the helium atom from its singlet ground
state to the first excited singlet state which, due to spin, has
multi-reference character. The optimized laser pulse which achieves
98.59\% of occupation in the excited state after a time of
250~a.u.\ is shown in Fig.~\ref{fig:helium_occupations}. We also show
the evolution of the first two natural occupation numbers which
starting close to one and zero, respectively, approach each other
during the propagation. In this situation, any description which
enforces constant occupation numbers will clearly not describe the
situation accurately.

To describe a situation where occupation numbers change in time,
functionals in RDMFT have to incorporate time-dependent approximations
for the cumulant of the reduced two-body density matrix
\cite{coleman-1963}. Possible approaches along these lines could be
based on reconstruction approaches for the cumulant of the two-body
matrix \cite{P2006}, or alternatively on antisymmetrized geminal
power (AGP) wave functions \cite{coleman-1965}, which was proposed
recently \cite{AG2010} and will be investigated in a future study.

\section{Conclusions and outlook}

In this work we aimed to partially unveil the role of
electron-correlation in the electron dynamics of systems driven out of
equilibrium. To reach this goal, we have used different 1D model
systems where we could assess, by comparing with the exact solution,
the quality of density and reduced density-matrix functionals in
various situations where the system interacts with external
time-dependent fields. We looked at both linear and non-linear
responses. The 1D ALDA approximation shows the same behavior as its
3D counterpart leading to the well-known underestimation of the
ionization potential and a failure to reproduce the excitations to
Rydberg states. Also, as has been seen in the past
\cite{MZCB2004,TK2009}, we showed that in the linear response function
double excitations do not appear when using adiabatic approximations
(exemplified here by using both ALDA and EXX).  For the case studied
here, where spatial symmetry has been broken and the ground state is a
described by a doubly occupied KS orbital, double excitations become
visible beyond linear response for the above mentioned adiabatic
functionals.  In going to the non-linear regime, we demonstrate that
the description of Rabi oscillations within all adiabatic functionals
leads to a dynamical detuning as the system is driven out of resonance
by the changes in the potential due to the changing density associated
with the transition during the Rabi oscillation. This manifests itself
in a very small population of the excited state in contrast to the
exact resonant propagation.  Hence, the description of Rabi
oscillations provides a very good test case for the development of
non-adiabatic functionals.


Within RDMFT adiabatic extensions of commonly employed ground-state
functionals lead to constant occupation numbers.  This was shown to be
a valid description within linear response but it turns out to be a
poor approximation in situations where transitions to and among
excited states (with possible multi-reference character) take place
during the evolution of the system.

In the future, the 1D model systems will be used to improve existing
approximations, especially going beyond the adiabatic dependence on
the density or the density matrix. Possible routes along these lines
include e.g.~orbital functionals in TDDFT, or explicit cumulant
approximations in time-dependent RDMFT.

The authors thank Michele Casula, Matthieu Verstraete, Miguel Marques,
and Xavier Andrade for helpful discussions. We acknowledge support by
MICINN (FIS2010-21282-C02-01), ACI-promociona (ACI2009-1036), ''Grupos
Consolidados UPV/EHU del Gobierno Vasco'' (IT-319-07), and the
European Community through e-I3 ETSF project (Contract No. 211956).


\end{document}